\documentstyle[epsf]{article}
\newcommand{\bee}{\begin{equation}}
\newcommand{\ee}{\end{equation}}
\newcommand{\beea}{\begin{eqnarray}}
\newcommand{\eea}{\end{eqnarray}}

\begin{document}
\thispagestyle{empty}
\parskip=12pt
\raggedbottom

\def\mytoday#1{{ } \ifcase\month \or
 January\or February\or March\or April\or May\or June\or
 July\or August\or September\or October\or November\or December\fi
 \space \number\year}
\noindent
\hspace*{9cm} COLO-HEP-370\\
\vspace*{1cm}
\begin{center}
{\LARGE Scaling and Topological Charge
 of a Fixed Point Action for $SU(2)$ Gauge Theory}
\footnote{Work supported in part by
NSF Grant PHY-9023257 
and U.~S. Department of Energy grant DE--FG02--92ER--40672}

\vspace{1cm}

Thomas DeGrand,
Anna Hasenfratz, and Decai Zhu\\
Department of Physics \\
 University of Colorado,
Boulder CO 80309-390

\vspace{1cm}

\mytoday \\ \vspace*{1cm}

\nopagebreak[4]

\begin{abstract}
We construct a few parameter approximate fixed point action for
SU(2) pure gauge theory and subject it to scaling tests, via Monte Carlo
simulation.
We measure the critical coupling for deconfinement for lattices
of temporal extent $N_t=2$, 3, 4, the
 torelon mass at fixed physical volume, and the
string tension (and heavy quark potential) from Wilson loops.
We calculate the topological susceptibility using inverse blocking
and show that it scales over the observed range of lattice spacings.
\end{abstract}

\end{center}
\eject


\section{Introduction}
 In QCD instantons may be responsible for
breaking axial symmetry and resolving the $U(1)$ problem \cite{U1}.
The relevant observable is the 
topological susceptibility $\chi$, defined as the infinite volume
 limit of
\bee
\chi_t = {{\langle Q^2 \rangle}\over V}
\ee
where $Q$ is the topological charge and $V$ the space time volume.
In QCD $\chi_t$ is a dimension-4 object with no weak coupling expansion,
and a calculation of $\chi_t$ in physical units in the continuum
requires nonperturbative techniques.
In the large-$N_c$ limit the mass of the $\eta'$ is related to the
topological susceptibility through
the Witten-Veneziano formula\cite{WV}
\bee
m^2_{\eta'} + m^2_{\eta} - 2m^2_K = 2 N_f \chi_t/f_\pi^2.
\label{WZF}
\ee

Unfortunately, the study of topology in lattice simulations is badly
contaminated by the presence of lattice artifacts, which
  can arise both from the form of the lattice action
and from the choice of lattice operator to define and measure
topological charge. 
A lattice action is, in general, not scale invariant, and so 
if a smooth continuum instanton is placed on the lattice, its action
can depend on its size. A lattice simulation can be compromised by
lattice artifacts, called ``dislocations,''
\cite{DISLOCATION},
which are non-zero charged configurations whose contribution to the
topological charge comes entirely from small localized regions.
If the minimal action of a dislocation is smaller than some critical
fraction (6/11 for SU(2)) 
of the continuum
value of a one-instanton configuration, then dislocations will dominate
the path integral and spoil the scaling of $\chi_t$ \cite{PANDT,BREAKDOWN}.
Difficulties also arise because the topological charge is not conserved on the
lattice. When the size of an instanton becomes small compared to the lattice
spacing it can "fall through" the lattice and its charge disappears. Topological
charge operators can fail to identify this process.

In Ref. \cite{INSTANTON1} we described a  framework which
solves both these problems and allows one to study topology in the
context of numerical lattice simulations in a theoretically reliable way.
If one can construct a lattice action and lattice operators
which live on the renormalized trajectory (RT) of some 
renormalization group transformation (RGT), then one's predictions
do not depend on the lattice spacing. A recent series of 
papers\cite{HN,PAPER1,PAPER2,PAPER3} have constructed
 fixed point (FP) actions for asymptotically free spin and gauge theories.
FP actions share the
scaling properties of the RT (through   one-loop quantum corrections)
and as such may be taken as a first approximation to a RT.
FP gauge actions have scale-invariant instanton solutions with an action value
of exactly $8 \pi^2/g^2$ and,   using RG techniques,
one can define a topological charge which has no lattice artifacts.
Basically, one inverts an RG transformation using a FP action
to produce a fine-grained lattice which has the same topological
structure as the original coarse configuration. On the fine lattice the size of any
topological object scales as the ratio of coarse to fine lattice spacings,
and after enough blocking steps the objects will become so large that
their charge can be measured in any way desired.

The purpose of this paper is to construct an approximate FP action
for SU(2) pure gauge theory,
subject it to scaling tests, and then measure the topological susceptibility.
In the next section we outline the construction of an action. 
 Then we describe its properties
under simulation: we have computed the critical temperatures for
deconfinement,  and the string tension measured using correlators of
Polyakov loops and from Wilson loops. The action shows good scaling
properties beginning at lattice spacing $a \le 1/2T_c$. Finally we
describe our measurement of the topological susceptibility using
inverse blocking.

We (again) remark that our approach to topology
has been anticipated by studies by others of 
 two-dimensional spin models\cite{INSTANTON,CP3,DEILIA},
although we cannot go as far due to computer speed and
memory limitations.

\section{Construction of an Approximate FP Action}

In order to have a practical action for use in simulations,
we adopt a slightly different method for finding a FP action than was used in
Ref. \cite{INSTANTON1}. We want an action valid only for  configurations
which would be encountered in a realistic simulation, with correlation lengths
of a few lattice spacings. Thus we solve the FP steepest descent
equation 
\bee
S^{FP}(V)=\min_{ \{U\} } \left( S^{FP}(U) +T(U,V)\right),  \label{STEEP}
\ee
(where
$\{V\}$ is the coarse configuration, $\{U\}$ is the fine configuration,
and $T(U,V)$ is the blocking transformation)
as follows: we generate coarse configurations using the Wilson action
with Wilson coupling $\beta=1.5-2.5$
 and perform the minimization to
find the number $S^{FP}(V)$. 
We use an algorithm similar to the one we employed for $SU(3)$
gauge theory \cite{PAPER2}, making a random rotation on each link variable and
quadratically interpolating to a proposed minimum.
The cost of a sweep through the lattice is  roughly equivalent
to a standard Monte Carlo update sweep.
We reduced the change in the action per sweep to a value of 0.005
or below. Typical action values range from 1000 to 8000, depending on the volume  of
the lattice and the coupling.
 The number of sweeps required to minimize the configuration to the given accuracy 
 is strongly beta-dependent, the procedure gets increasingly costly 
at small $\beta$ values.

We fit $S^{FP}(V)$ measured according to Eqn. \ref{STEEP} to a small set of
operators and couplings. The fit is not designed to work for
configurations which are much smoother or coarser
than the ones we present to be
minimized. 
There is considerable freedom in this fitting method, and it may be that
alternate approximate FP actions exist which solve Eqn. \ref{STEEP}
equally well, and have other nicer properties in addition.

\begin{figure}[htb]
\begin{center}
\vskip 10mm
\leavevmode
\epsfxsize=90mm
\epsfbox{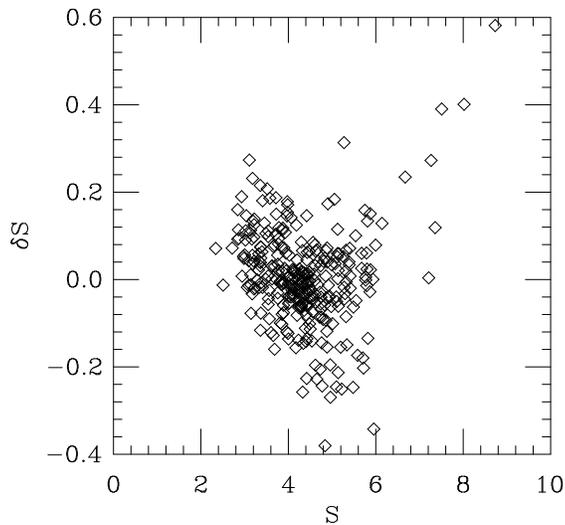}
\vskip 10mm
\end{center}
\caption{ Difference between the measured (Eqn. 3) and fitted (Eqn. 4) value of
$S_{FP}(V)$ for the
eight parameter approximate FP action of Table 1.}
\label{fig:deltas}
\end{figure}

 The approximate FP action consists of several powers of
two loops, the plaquette and the perimeter-six loop (x,y,z,-x,-y,-z),
\bee
S(V) = {1 \over 2} \sum_{\it C} ( c_1({\it C})(2-{\rm Tr}(V_{\it C})) +
                       c_2({\it C})(2-{\rm Tr}(V_{\it C}))^2+ ...
\label{ACTIONPAR}
\ee
with coefficients tabulated in Table \ref{tab:eightpar}.
The quality of the fit is shown in the scatter plot of Fig. \ref{fig:deltas}.
We chose the $c_1$ coefficients so that the 
approximate FP action 
is $O(a^2)$ improved.

\begin{table*}[hbt]
\caption{Couplings of the few-parameter FP action for SU(2) pure gauge theory.}
\label{tab:eightpar}
\begin{tabular*}{\textwidth}{@{}l@{\extracolsep{\fill}}lcccccc}
\hline
operator &  $c_1$ & $c_2$ & $c_3$ & $c_4$ \\
 \hline
$c_{plaq}$ &  .3333  &  .00402      & .00674 & .0152   \\
$c_{6-link}$ &  .08333  &  .0156 &  .0149 & -.0035        \\
\hline
\end{tabular*}
\end{table*}

\begin{figure}[htb]
\begin{center}
\vskip 10mm
\leavevmode
\epsfxsize=90mm
\epsfbox{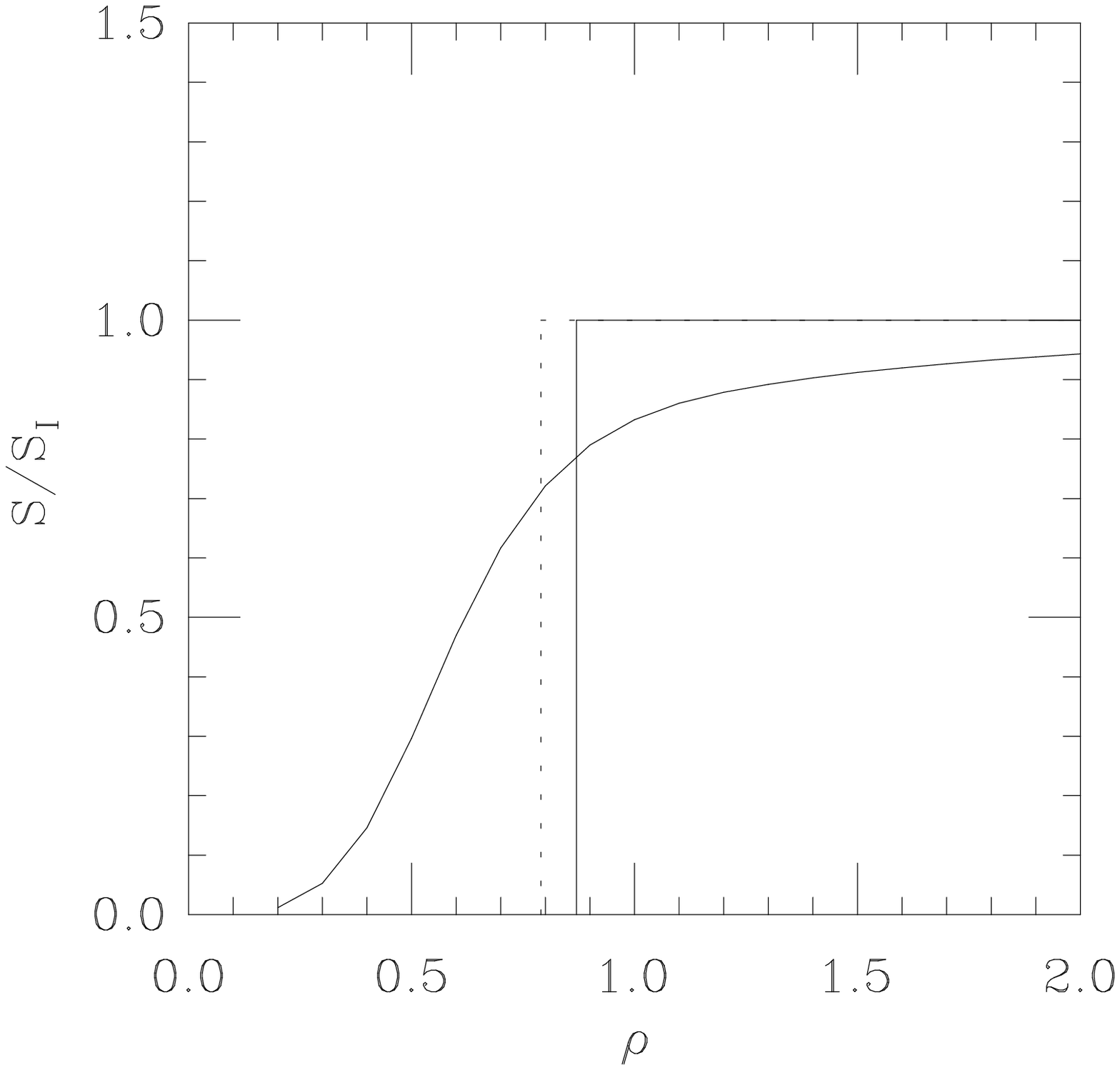}
\vskip 10mm
\end{center}
\caption{ Action of single blocked $c_1$
instanton configurations computed using the eight parameter FP action.
The dotted line indicates the
$Q=0 \to Q=1$ boundary on the original coarse lattice while the solid line is the
boundary on the inverse blocked configuration.}
\label{fig:sbc1}
\end{figure}

\begin{figure}[htb]
\begin{center}
\vskip 10mm
\leavevmode
\epsfxsize=90mm
\epsfbox{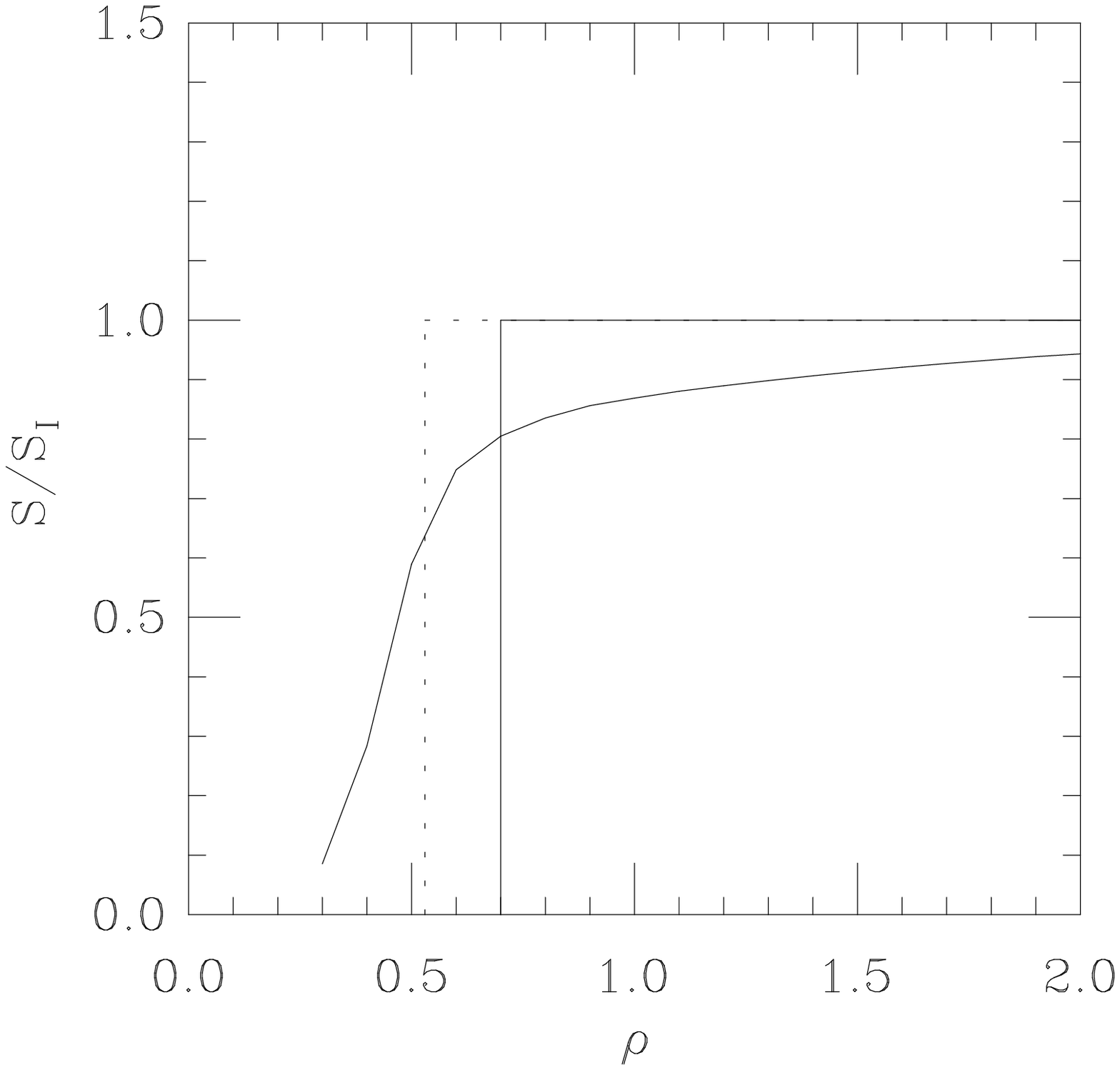}
\vskip 10mm
\end{center}
\caption{ Action of single blocked $c_2$
instanton configurations computed using the eight parameter FP action,
displayed as in Fig. 2.}
\label{fig:sbc28}
\end{figure}

In Ref. \cite{INSTANTON1} we constructed sets of trial smooth instantons
and measured their action and charge. We found that when they had
$Q=1$, their action 
$S(\rho) \ge S_I$,  where $S_I$ is
the continuum instanton action, and that dislocations
(instanton-like field configurations for which $Q \ne 1$) have actions
evaluated at the FP using
 inverse blocking which is less than $S_I$.
Now we make a similar comparison for these configurations using our
approximate FP action.  We consider two kinds of smooth
configurations, singly blocked $c_1$ and $c_2$ instantons (for
a description of these objects see Ref. \cite{INSTANTON1}.)
We show actions versus instanton size $\rho$ in Figs. \ref{fig:sbc1}
and \ref{fig:sbc28}.
The dotted line indicates the
$Q=0 \to Q=1$ boundary on the original coarse lattice while the solid line
 is the boundary on the inverse blocked configuration. 
Instanton configurations were not included in the fit which
 produced the action. 
The curve shows the action on the coarse lattice evaluated according to
Eqn. \ref{STEEP} and Table \ref{tab:eightpar}.
The action does not reproduce the good feature of
the full FP action: in particular, $S(\rho)<S_I$ out to very large $\rho$.
However, this action is still quite suitable for simulations. It will
generate dislocations (configurations which have $Q=1$ but
$S<S_I$), but those dislocations will not dominate the functional integral
because their action exceeds the entropic bound (for SU(N)) of
\bee
S > {{48 \pi^2} \over {11N^2}},
\ee
or $S/S_I > 6/11$ for $SU(2)$.
If we measure the action on an inverse blocked lattice using
a parameterization of the FP action appropriate to smooth gauge configurations
we will reproduce
the results of Ref. \cite{INSTANTON1}, namely $S(\rho) \ge S_I$ when $Q=1$,
$Q=0$ when $S(\rho)<S_I$.

\section{Critical Couplings for Deconfinement}

The deconfinement phase transition for the pure  SU(2) gauge theory is  second
order,  and the transition
 is in the same universality class as the three-dimensional Ising
model. The order parameter is the average Polyakov loop
\begin{eqnarray}
L & = &  \frac{1}{N_{s}^3} \sum_{\vec{x}} L(\vec{x} )    
\end{eqnarray}
where 
\begin{eqnarray}
L(\vec{x}) & = & \frac{1}{2} Tr \prod_{t=1}^{t=N_{t}} U_{t,\vec{x};0}
\end{eqnarray}
and
$N_{s}$ and $N_{t}$ are  the  spatial and temporal size
of the lattice respectively. 

The  Binder cumulant is defined as
\begin{eqnarray}
g_{4} &= & \frac{<L^4>}{<L^2>^2} -3 .
\end{eqnarray}
A standard finite size scaling analysis shows that this quantity is a scaling
function \cite{ENGELS}.
It must be independent of $N_{s}$ at the critical point (neglecting 
contributions from irrelevant scaling fields).

In order to determine the critical coupling,  
we run simulations at different $N_{s}$ for each $N_{t}$, then
plot the Binder cumulant. The location in $\beta$ at which
the different curves cross must correspond to the
critical point.

\begin{figure}[htb]
\begin{center}
\vskip 10mm
\leavevmode
\epsfxsize=90mm
\epsfbox{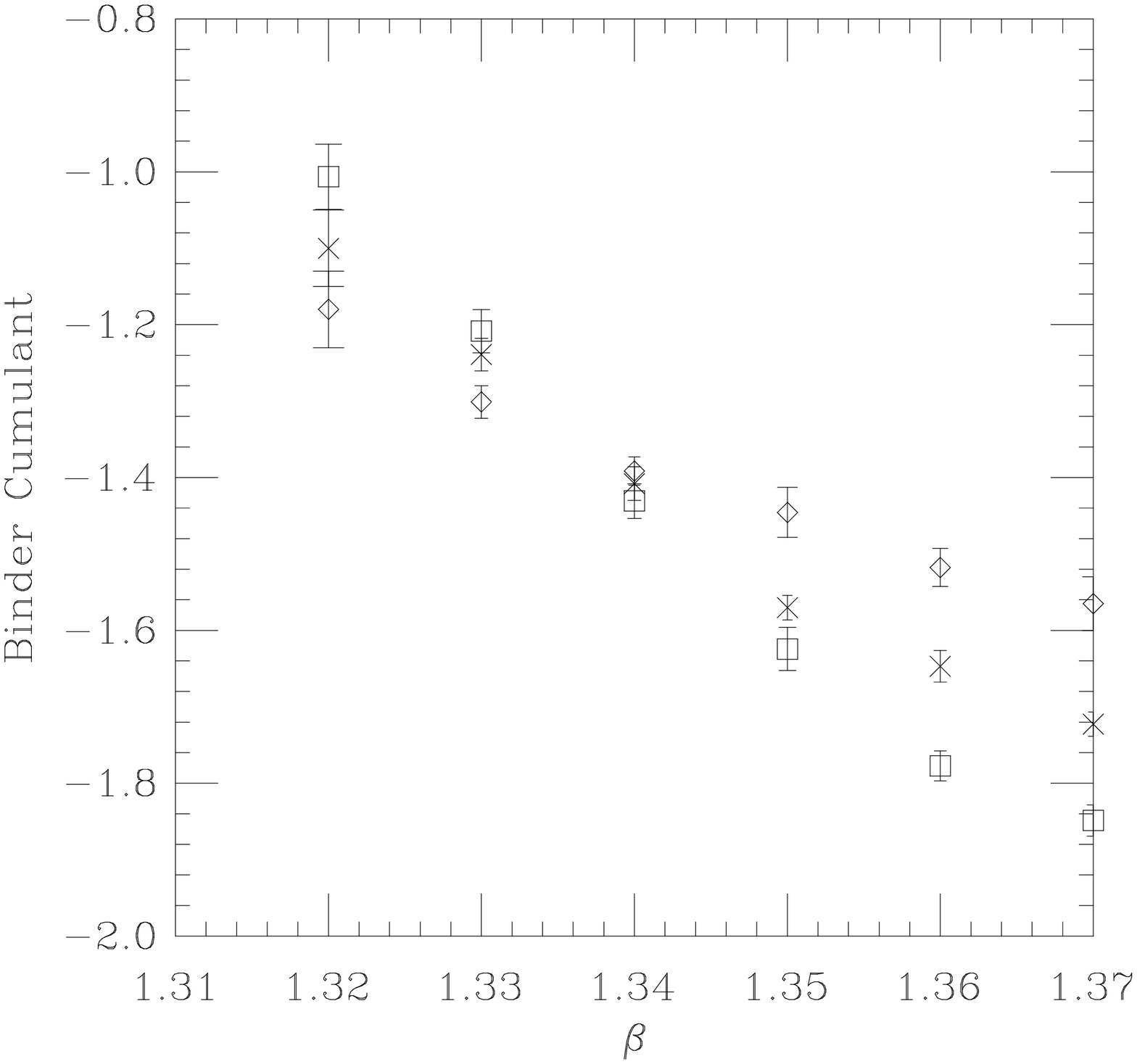}
\vskip 10mm
\end{center}
\caption{ The Binder cumulant for the FP action for $N_{t}=2$. The
diamond shows results for $4^3$ spatial volumes,  the
cross labels $6^3$ volumes, and the square labels $8^3$.
}
\label{fig:bcumulant_nt2}
\end{figure}

Fig. \ref{fig:bcumulant_nt2} shows the Binder cumulant plot for
$N_{t} =2$. We performed simulations at three different volumes,  
$N_{s} =4, 6$ and 8. From this plot, we see that the crossing
happens  at  approximately $\beta=1.34$ and $g_{4} = -1.40$, which is 
consistent with the  result from the
three-dimensional Ising model where $g_{4}= -1.41(1)$
\cite{ISING}. 
Combining this plot and the requirement that
$g_{4} = -1.40$ at the crossing, we conclude that 
$\beta_{c} = 1.340(5)$ for $N_{t} = 2$.

\begin{figure}[htb]
\begin{center}
\vskip 10mm
\leavevmode
\epsfxsize=90mm
\epsfbox{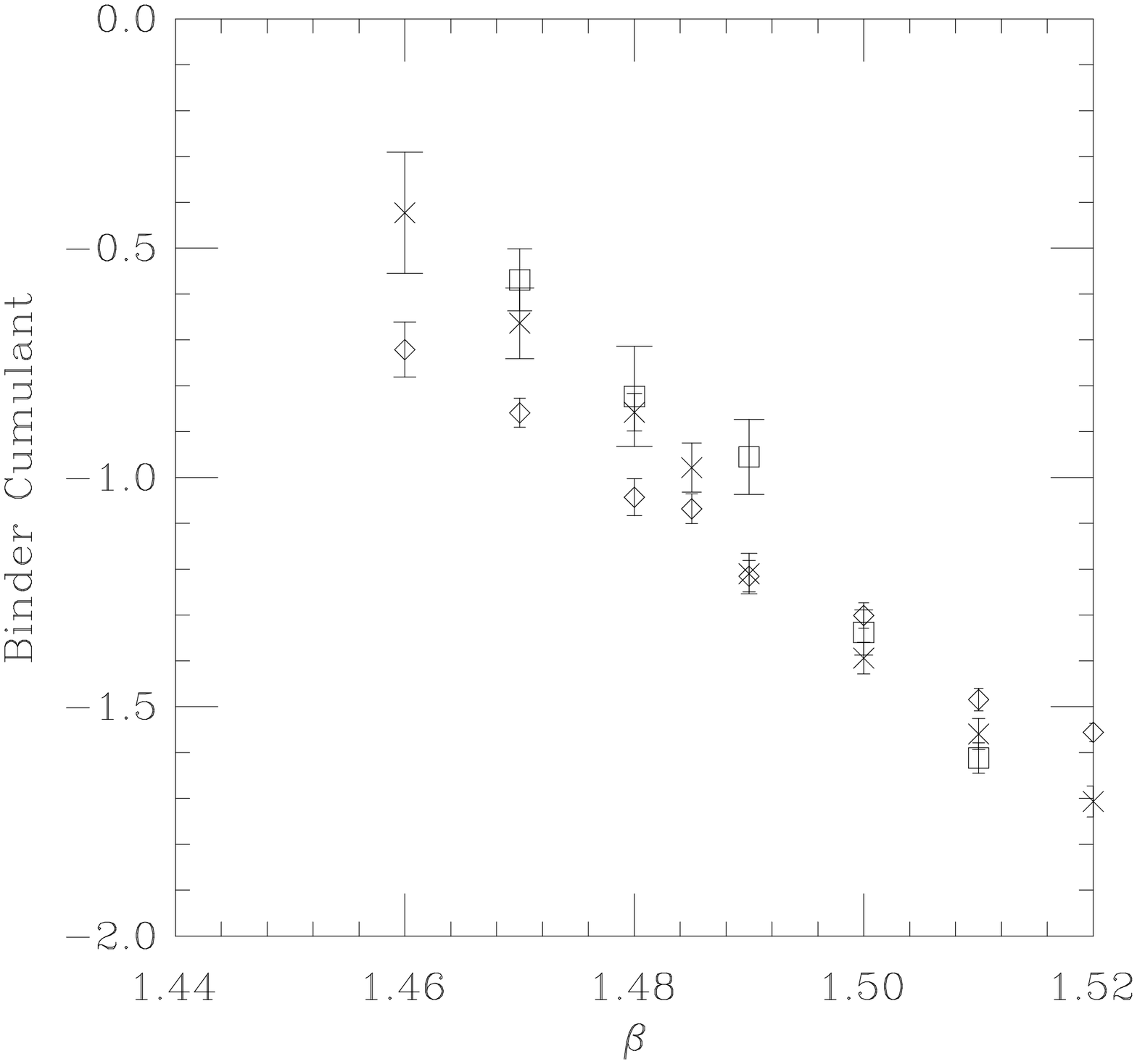}
\vskip 10mm
\end{center}
\caption{ The Binder cumulant for the FP action for $N_{t}=3$.
	Diamond: $6^4$, cross: $8^3$ and square: $10^3$ spatial volumes.  
}
\label{fig:bcumulant_nt3}
\end{figure}

\begin{figure}[htb]
\begin{center}
\vskip 10mm
\leavevmode
\epsfxsize=90mm
\epsfbox{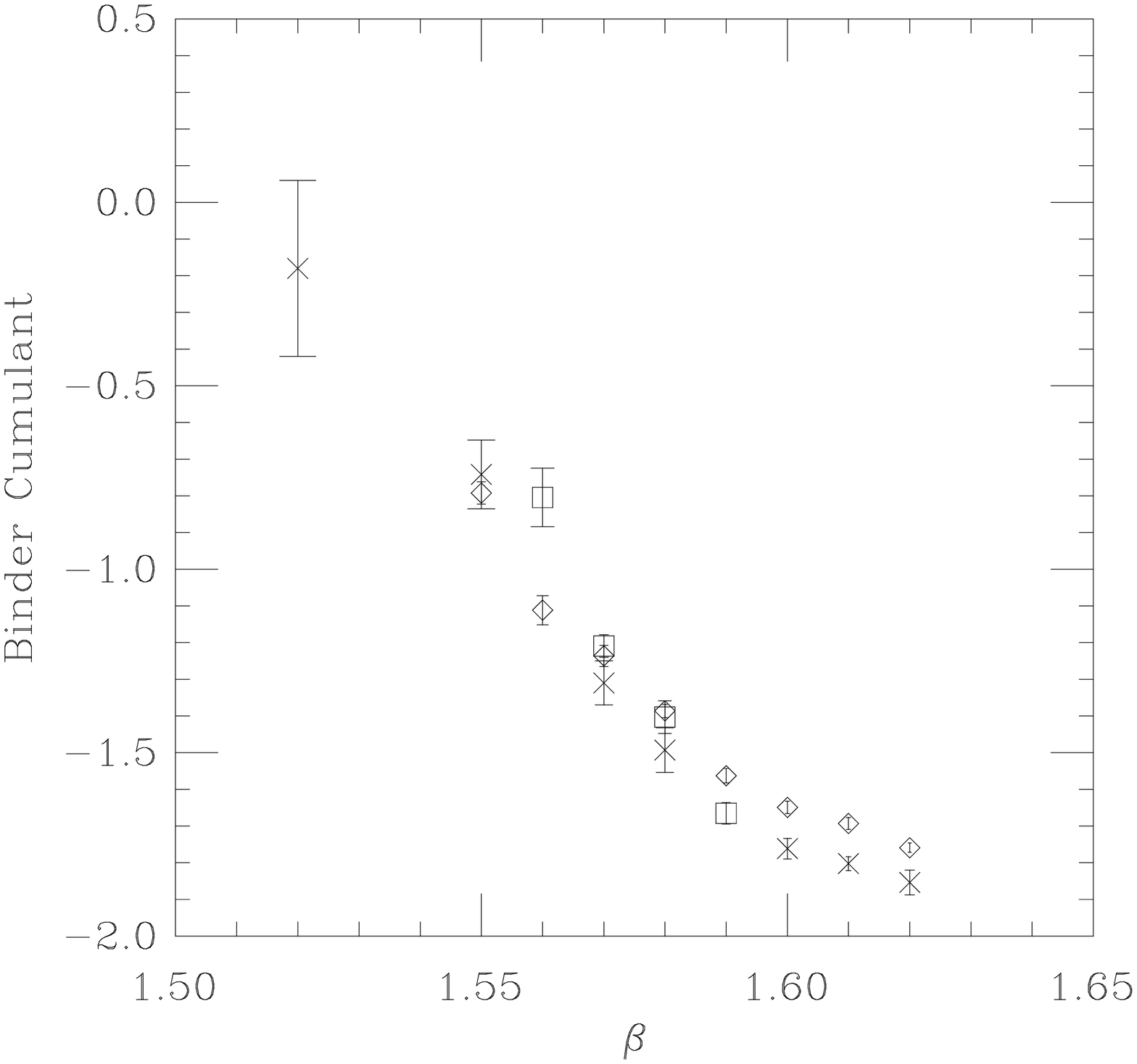}
\vskip 10mm
\end{center}
\caption{ The Binder cumulant for the FP action for $N_{t}=4$.
Diamond: $8^3$, cross: $10^3$, square: $12^3$ spatial volumes. 
}
\label{fig:bcumulant_nt4}
\end{figure}

Figs. \ref{fig:bcumulant_nt3} and \ref{fig:bcumulant_nt4} show 
the  Binder cumulant plots for $N_{t}=3$ and $N_{t} = 4$. It
is difficult to decrease the statistical error for  these data points. 
We identify the critical coupling as the point where
  $ g_{4}= -1.40$,
and so we conclude that $\beta_{c} = 1.502(5)$ and  1.575(10) for 
$N_{t}= 3$ and 4, respectively. 
 
\begin{figure}[htb]
\begin{center}
\vskip 10mm
\leavevmode
\epsfxsize=90mm
\epsfbox{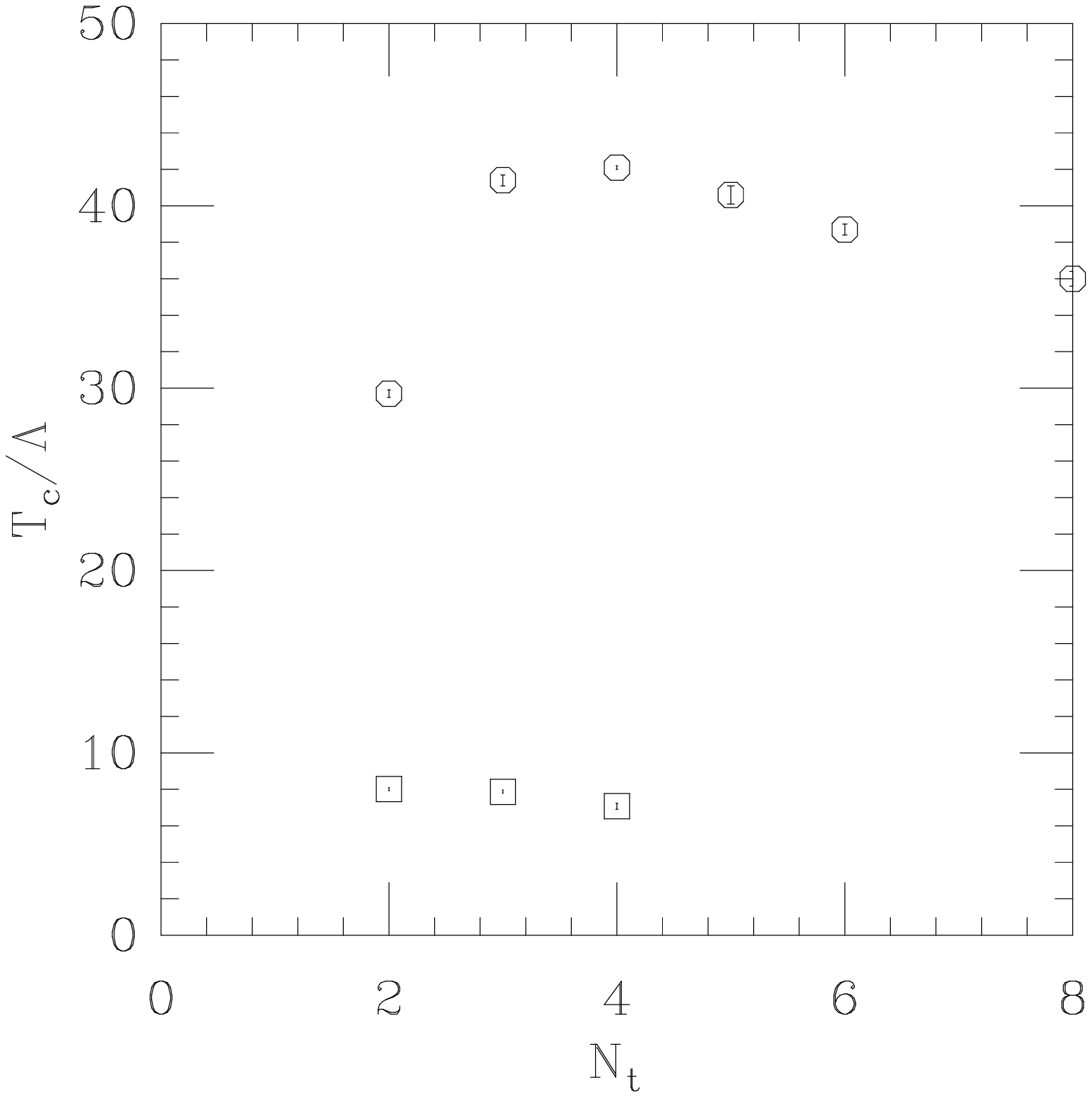}
\vskip 10mm
\end{center}
\caption{$T_c/\Lambda$ for the Wilson (octagons) and  FP (squares)
actions.
}
\label{fig:tc}
\end{figure}

Although the FP action is not designed to improve asymptotic scaling,
only scaling, we can still see if it shows asymptotic scaling.
Using the original coupling in the action, we compute $T_c/\Lambda$
from the two loop formula from our data, and record the results in Table
\ref{tab:betacrit} and Fig. \ref{fig:tc}. The eight parameter
FP action   shows asymptotic scaling within fifteen per cent
from $N_T=2$ to 4 in terms of the bare coupling, with $T_c/\Lambda \simeq
7-8$.
Its
$\Lambda$ parameter is about a factor of five larger than the Wilson
one and therefore much closer in value to the continuum $\Lambda$
parameters.

\begin{table*}[hbt]
\setlength{\tabcolsep}{1.5pc}
\caption{Critical couplings at 
infinite volume for the FP action with parameters in Table~1.}
\label{tab:betacrit}
\begin{tabular*}{\textwidth}{@{}l@{\extracolsep{\fill}}cccc}
\hline
$N_{t}$     & 2   & 3 & 4  \\
$\beta_{c}$ & 1.340(5) & 1.502(5) & 1.575(10)   \\
$T_c/\Lambda$   & 8.01(9) & 7.87(9) & 7.04(17)  \\
\hline
\end{tabular*}
\end{table*}

\section{Torelon Mass}

In this section we describe the measurement of the
 string tension $\sigma$  through the correlator of pairs of
objects with the quantum numbers of Polyakov loops: on a lattice
of transverse size $L$ the correlator of two Polyakov loops averaged
over transverse separations and separated
a longitudinal distance $z$ is
\begin{equation}
C(z) = \sum_{r_\perp} {\rm Re} \langle P_j(r_\perp,z)P_j^\dagger(0,0) \rangle
 \simeq \exp(-\mu z)
\end{equation}
(plus boundary terms)
where $\mu$ is the so-called torelon mass. On an infinite lattice
$\mu = \sigma L$ and we will define $\mu = L \sigma(L)$ on a finite lattice.
To carry out a scaling test we fix the lattice
spacing through the deconfinement temperature and
perform a series of simulations on lattices of fixed physical size $L= c/T_c$
and different lattice spacings. Here we take $c=2$ and $aT_c = 1/2$, 1/3,
and 1/4 (the corresponding couplings were determined in the previous section).
As a scaling test we compute $L\sqrt{\sigma}= c\sqrt{\sigma}/T_c$.

\begin{figure}[htb]
\begin{center}
\leavevmode
\epsfxsize=90mm
\epsfbox{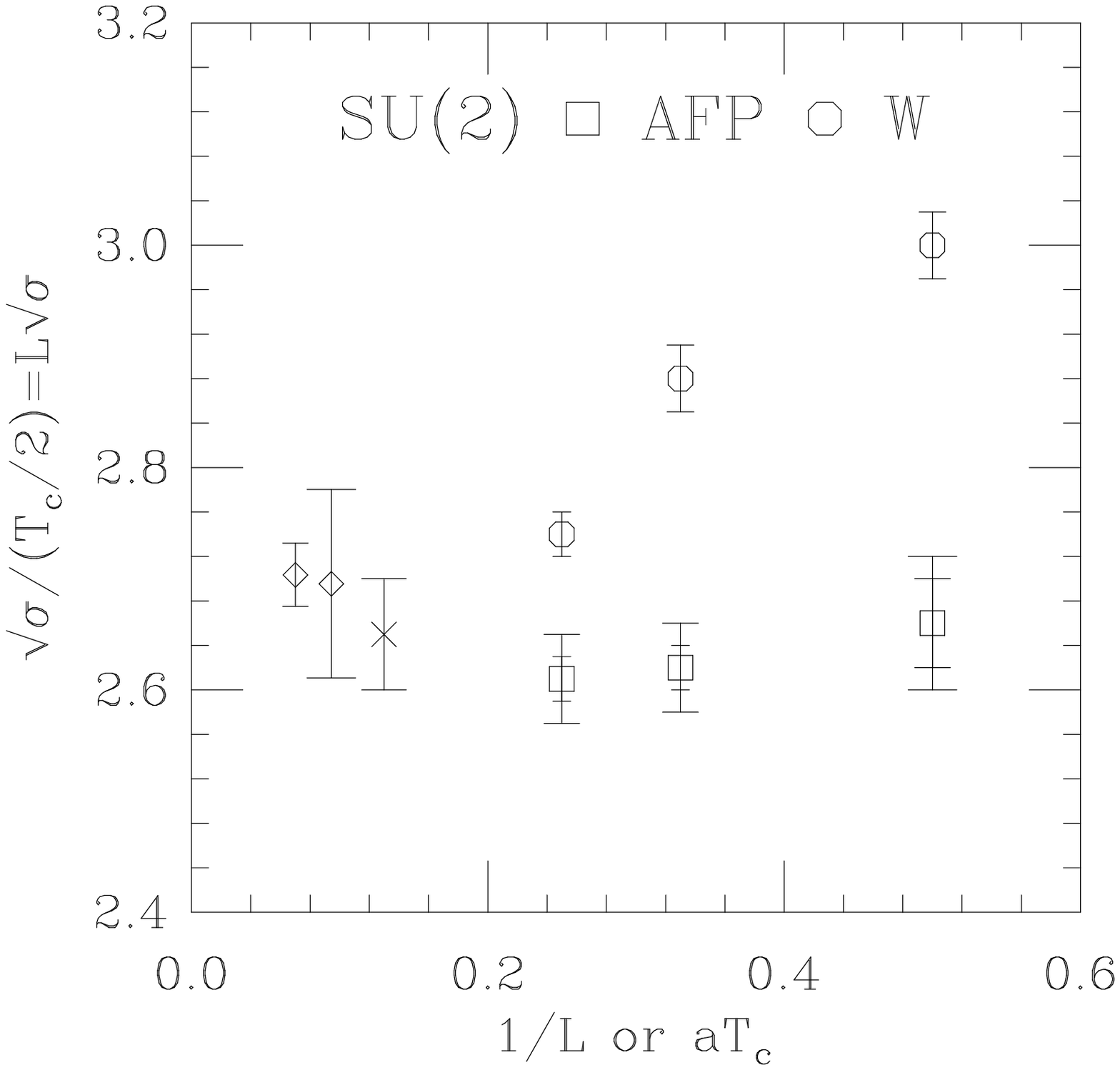}
\end{center}
\caption{Scaling test of torelon mass on lattices of fixed physical volume
for the Wilson action (crosses) and FP action
 (squares); $T_c$ is defined in infinite
spatial volumes.
The cross is a torelon mass measurement at $L=2.07/T_c$ from Ref. 18
and the diamonds are extrapolations by us to $L=2/T_c$.
}
\label{fig:aspect_2a}
\end{figure}

At large lattice spacing ($aT_c = 1/2$) the best signals came
from the Polyakov loop itself,
measured with the Parisi -
Petronzio - Rapuano \cite{PPR} multihit variance reduction method.
At smaller lattice spacing ($aT_c \le 1/3$) we used
 correlators of APE-blocked
\cite{APEBlock} links: we iterate
\begin{eqnarray}
V^{n+1}_j(x) = (1-\alpha)V^{n}_j(x) & +  & \alpha/4 \sum_{k \ne j}
(V^{n}_k(x)V^{n}_j(x+\hat k)V^{n}_j(x+\hat j)^\dagger
\nonumber  \\
& + & V^{n}_k(x- \hat k)^\dagger
 V^{n}_j(x- \hat k)V^{n}_j(x - \hat k +\hat j) )
\end{eqnarray}
(with $V^0_j(x)=U_j(x)$ and $V^{n+1}_j(x)$ projected back onto $SU(2)$),
with $\alpha$ varying from 0.2 at $aT_c=1/3$ to
0.5 at $aT_c = 1/4$ 
and ten blocking steps.
  
We determined the torelon mass from a single-exponential correlated fit,
beginning at a minimum $z$ where the $\chi^2/DF$ is near unity, and where we
see stability in the effective mass over a range of $z$. At $aT_c = 1/2$
the fits are to $z=1$ and $z=2$ points only since there is no signal for $z>2$.

We display our results in Fig. \ref{fig:aspect_2a} and Tables
\ref{tab:tor_w} and \ref{tab:tor_afp}.
 An uncertainty in the value of the critical coupling  propagates into
the uncertainty of $L\sqrt{\sigma(L)}= 2 \sqrt{\sigma}/T_c$ for the FP action.
 We include this uncertainty in the figure
by combining it in quadrature with the statistical fluctuation 
in $L\sqrt{\sigma(L)}$;
in the figure the extreme range of the vertical error bars shows
the combined uncertainty.

\begin{table*}[hbt]
\setlength{\tabcolsep}{1.5pc}
\caption{Our measurements of torelon masses from the Wilson action on
small lattices.}
\label{tab:tor_w}
\begin{tabular*}{\textwidth}{@{}l@{\extracolsep{\fill}}lcccccc}
\hline
volume &  $\beta$ &  $\mu = L \sigma$ & $L\sqrt{\sigma}$\\
 \hline
 \hline
$4^3 \times 12$  & 1.88  & 2.25(5) & 3.00(3) \\
\hline
$6^3 \times 12$  & 2.177  & 1.38(3) & 2.88(3)  \\
\hline
$8^3 \times 12$  & 2.29872  & 0.97(3) & 2.78(4) \\
\hline
\end{tabular*}
\end{table*}
 
\begin{table*}[hbt]
\setlength{\tabcolsep}{1.5pc}
\caption{Our measurements of torelon masses from the FP action.}
\label{tab:tor_afp}
\begin{tabular*}{\textwidth}{@{}l@{\extracolsep{\fill}}lcccccc}
\hline
volume &  $\beta$  & $\mu = L \sigma$ & $L\sqrt{\sigma}$\\
 \hline
 \hline
$4^3 \times 12$  & 1.34   & 1.75(6) & 2.65(4) \\
\hline
$6^3 \times 12$  & 1.502   & 1.16(7) & 2.62(2) \\
\hline
$8^3 \times 12$  & 1.575   & 0.82(3) & 2.55(4) \\
\hline
\end{tabular*}
\end{table*}

There have been many simulations in the literature which
present the string tension from the torelon mass \cite{MANDT,MANDP,SB}.
However, most of them are done on lattices whose physical size is not
the one used here, $L=2/T_c$. On the scale of Fig. \ref{fig:aspect_2a}
the torelon mass $\sigma(L)$ is quite sensitive to $L$.
We elect to show only three small lattice spacing points.
The first is the Wilson action $\beta=2.50$ $16^4$ result
of Michael and Teper\cite{MANDT}; for this data set, $L \simeq 2.07/T_c$
(a lattice with $N_t=7.7$ would be at its deconfinement transition
at $\beta=2.50$). We also show the $\beta=2.60$ point of the same authors 
and the $\beta=2.70$ point of Ref. \cite{MANDP}: for these data we
extrapolate to $L=2/T_c$ using the string formula
\bee
\sigma = \sigma(L) + {{\pi}\over{3L^2}}
\ee
(which itself may not be correct).
 
The result of our simulations is that the torelon mass measured
on aspect ratio 2 lattices using the FP action scales within one
standard deviation
for $1/4 \le aT_c \le 1/2$, at a value which appears to be consistent with
the value inferred from Wilson action results from small lattice
spacing simulations.
 
\section{Potential from Wilson Loops}
As $\beta$ rises, the critical temperature for deconfinement becomes
more difficult to measure, and one needs some other quantity to determine
the scale. Torelons are not a good choice because of their sensitivity
to volume. The only other relatively straightforward observable is the
potential, extracted from $L \times T$ Wilson loops $W(L,T)$.

We used a method (and a program) developed by U.~Heller\cite{HELV}: we construct
APE-blocked spatial links and measure Wilson loops made of these
blocked spatial links and unblocked temporal links. This gives
a better overlap onto the potential.
To speed up the program,
the computation of the Wilson loops is done by first performing a gauge
transformation to axial gauge, after which
the desired APE smearing of space-like links is accomplished.
We measure all on-axis loops, as well as
 off-axis time-like loops with
the space-like parts along the directions (1,1,0), (2,1,0) and (1,1,1)
plus others related by the lattice symmetry.

The potential is defined up to an additive constant by
\bee
\exp(-V(R)T) = W(R,T)
\ee
and is
measured by fitting the analog of the ``effective mass'' in spectroscopy:
look at the ratio $-\log(W(R,T+1)/W(R,T))$ and ask for it to plateau;
then equate it to the potential $V(R)$.

These are low-statistics simulations. We used a $12^3 \times 16$ lattice
($8^4$ for Wilson $\beta=1.88$, FP $\beta\le 1.45$)
and collected 100-500 measurements of Wilson loops (spaced five update sweeps
apart) per coupling.

While the numerics are easy the analysis is not. The signal from big
$R$ disappears into the noise exponentially in $R$ (as the torelon mass
does) and the extraction of $V(R)$ from the data requires a global fit to
some functional form. However, as far as we can tell, the choice of
a particular ansatz is arbitrary. It is difficult to make reliable
error estimates if the results depend on it.

In Table \ref{tab:potential}
 we present the result of a fit to $V(r)$ to the form
\bee
V(r) = V_0 + \sigma r - E/r - F(G_L(r)-1/r)
\ee
with $G_L(r)$ the lattice Coulomb potential for the quadratic
action (i.e. the action given by truncating the formula of Table 1
by keeping only the $c_1$ terms, then expanding $2- {\rm Tr}U$ to
quadratic order in the vector potential).
A fit to the form 
\bee
V(r) = V_0 + \sigma r - E/r
\ee
produces essentially identical results for $\sigma$ for the smaller
lattice spacing data ($\beta>1.502$). The larger lattice spacing data 
are difficult to fit. The FP $\beta=1.34$ data set is fit keeping only
on-axis points, and is fit to pure linear plus Coulomb form. The Wilson 
$\beta=1.88$ data is fit keeping only the linear term.

A scale related to the string tension, which has smaller dependence
on the functional form of $V(r)$, is the distance $r_0$,
as defined through the force, by $r_0^2F(r_0)= -1.65$ \cite{SOMMER}.
In the physical world of three colors and four flavors, $r_0 = 0.5$ fm.
The last column of Table \ref{tab:potential} shows $r_0$ for our data.

\begin{table*}[hbt]
\setlength{\tabcolsep}{1.5pc}
\caption{Parameters of the static potential from Wilson loops.}
\label{tab:potential}
\begin{tabular*}{\textwidth}{@{}l@{\extracolsep{\fill}}lcccccc}
\hline
FP action \\
\hline
$\beta$   & $a^2\sigma$   & $\chi^2$ & $r_{min}$ & $r_{max}$ & $r_0/a$\\
\hline
 1.7  &  0.051(1) &  12.5  & 1.00  &   7.07 &  5.10(15) \\
 1.65  & 0.082(1) &   6.7 &  1.41  &   6.93 &  4.15(10) \\
 1.625 & 0.093(2) &   8.1  & 2.00  &   8.49 & 3.9(1)  \\
 1.6   & 0.111(2)  & 17.138  &  1.73  &   8.66 & 3.55(10)  \\
 1.575 & 0.133(2) &   14.5  &  1.73   &  10.3 &  3.25(10) \\
 1.502 & 0.237(7) &  14.5  &   2.24  &   8.49 &  2.48(5) \\
 1.45  & 0.35(1) &  5.6  &   2.24  &   5.56 &  2.05(10) \\
 1.34  & 0.51(2) &  1.0  &   1.00  &   4.00 &  1.7(1) \\
\hline
Wilson action \\
2.2987  &  .142(2)& 3.2 & 2.24 & 5.2 & 3.1(1)   \\
1.88   &  0.67(2) & .07& 2& 4 &  1.6(1) \\
\hline
\end{tabular*}
\end{table*}

The $\beta=1.34$, 1.502,
 and 1.575 points can be used for a second scaling test
since they are at $aT_c=1/2$, 1/3 and 1/4: $\sqrt{\sigma}/T_c$= 1.43(3),
1.46(2), and  1.46(1).
(The errors do not include an uncertainty in $\beta_c$). 

In contrast the Wilson action string tension
 shows an eight per cent violation of scaling between $N_t=2$ and 4, 
with $\sqrt{\sigma}/T_c$ at $N_t=2$ of 1.63(3) and at $N_t=4$ of 1.51(1).

\begin{figure}[htb]
\begin{center}
\leavevmode
\epsfxsize=90mm
\epsfbox{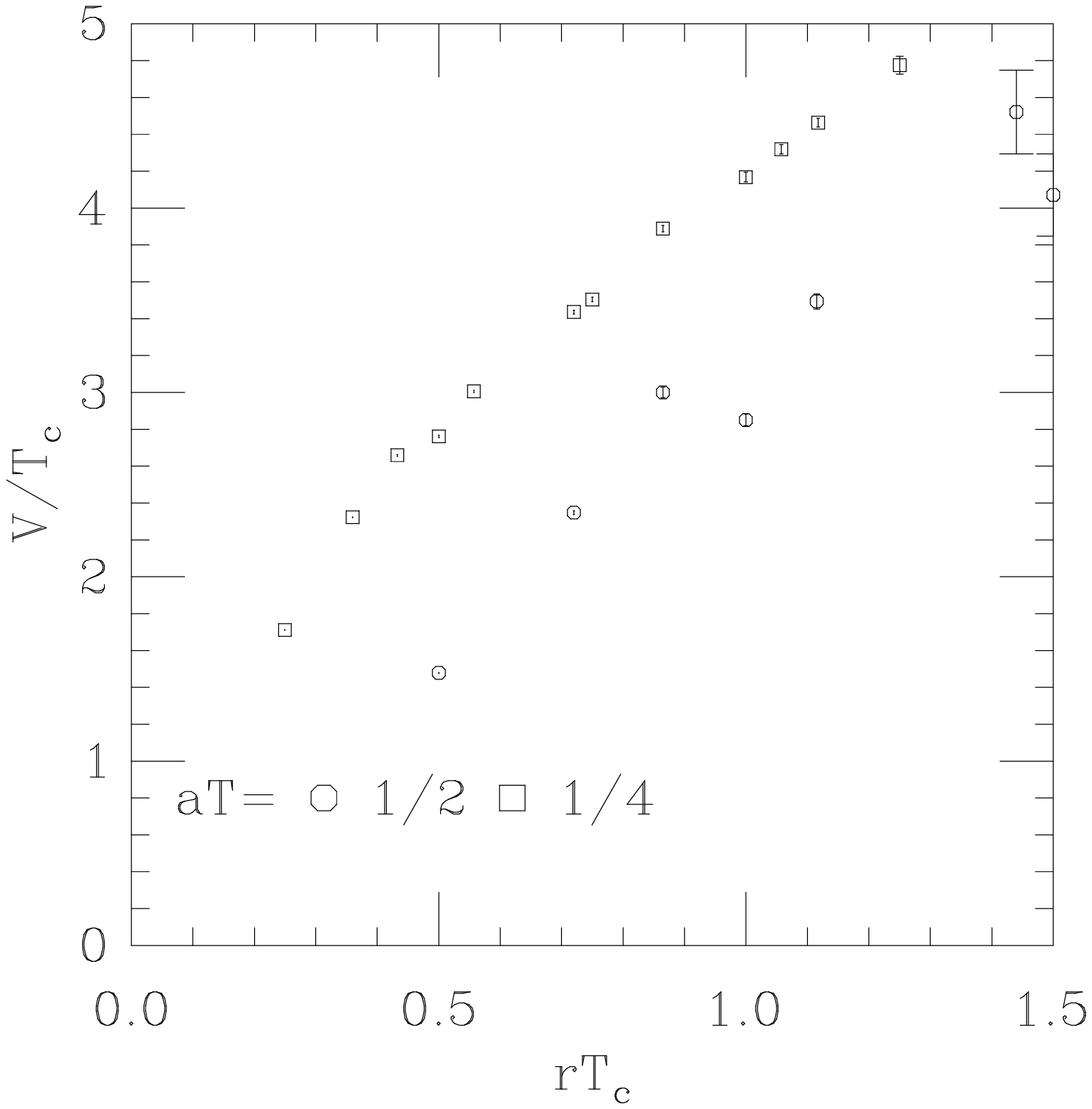}
\end{center}
\caption{  Potential $V(r)/T_c$ vs. $rT_c$ for the Wilson action at
$\beta_c(N_T=2)$ (octagons) and $\beta_c(N_T=4)$ (squares).  }
\label{fig:wilson24}
\end{figure}

\begin{figure}[htb]
\begin{center}
\leavevmode
\epsfxsize=90mm
\epsfbox{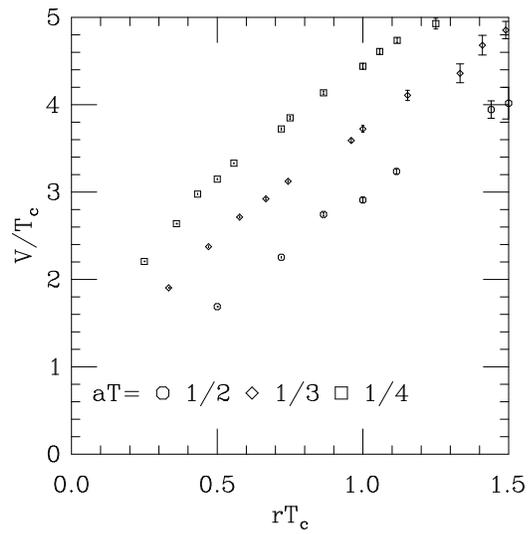}
\end{center}
\caption{ Potential $V(r)/T_c$ vs. $rT_c$ for the FP action at
$\beta_c(N_T=2)$ (octagons), $\beta_c(N_T=3)$ (diamonds)
 and $\beta_c(N_T=4)$ (squares).}
\label{fig:perfect24}
\end{figure}

Because we only use naive operators for the potential, not FP
operators\cite{PAPER1}, the presence or absence of rotational symmetry
breaking is not a scaling test. The persistence of such violation as the
lattice spacing changes is a test of scale breaking.
We can compare violations of rotational symmetry as a function of
lattice spacing by using our results from lattice spacings determined
from $T_c$, by scaling the potential by $1/T_c$ and the radius by $T_c$.
We display the scaled potentials from the Wilson action
 in Fig.~\ref{fig:wilson24}
and from the FP action in Fig.~\ref{fig:perfect24}.
We have not forced the potentials to lie atop each other, so the reader can
view the extent of rotational symmetry violations.

\begin{figure}[htb]
\begin{center}
\vskip 10mm
\leavevmode
\epsfxsize=90mm
\epsfbox{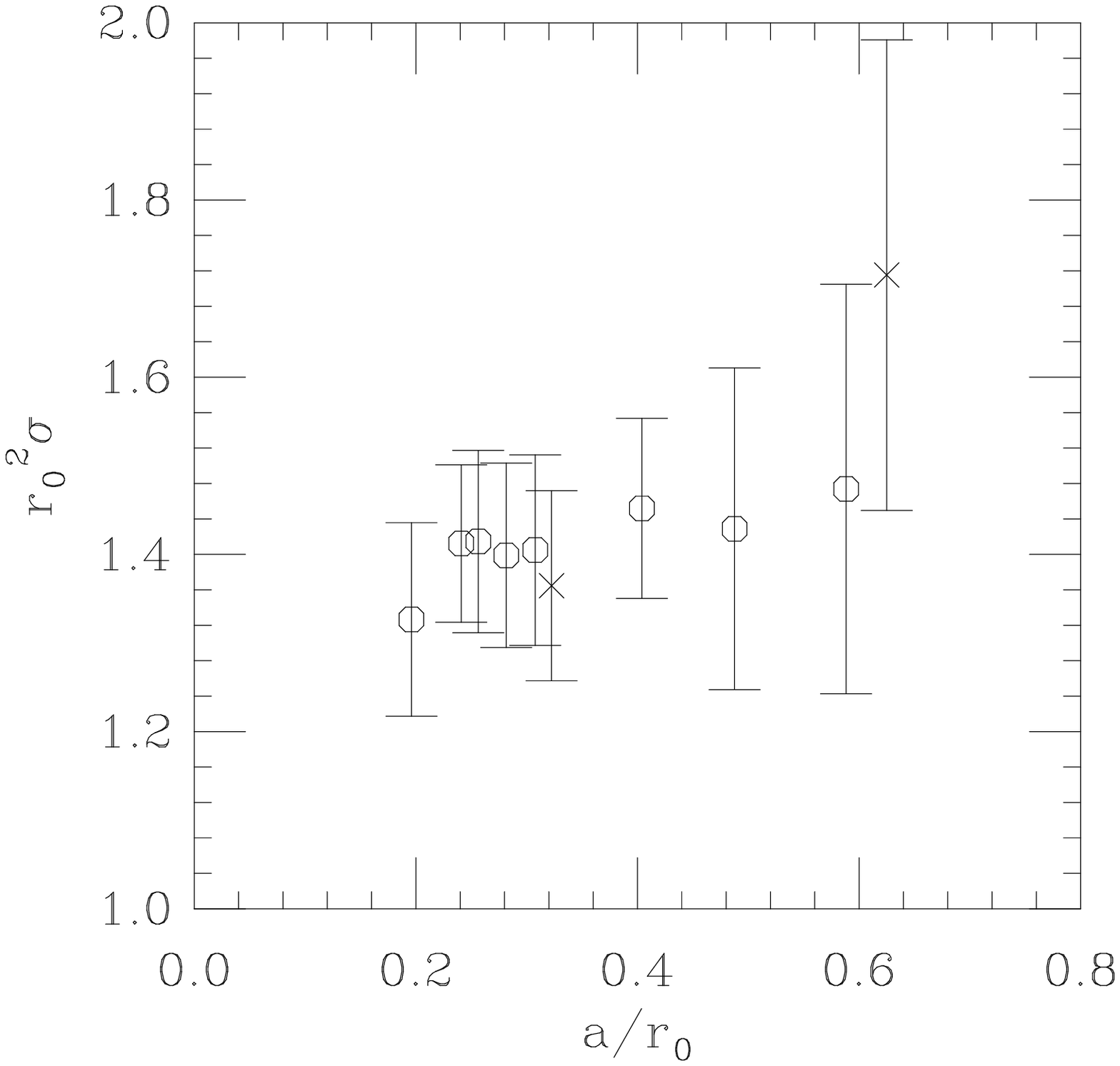}
\vskip 10mm
\end{center}
\caption{ String tension times $r_0^2$ for the FP action (octagons)
and Wilson action (crosses) as a function of $r_0/a$.}
\label{fig:sigmar0}
\end{figure}

The combination $r_0^2\sigma$ is a dimensionless variable
whose variation with lattice spacing gives a scaling test (to the extent that the
determination of $r_0$ and/or $\sigma$ can be compromised by lack of
rotational invariance). We present a plot of $r_0^2\sigma$ vs $a/r_0$
(lattice spacing in units of 0.5 fm) in Fig. \ref{fig:sigmar0}.
The dominant uncertainty is from the string tension. Both Wilson and
FP actions scale within our uncertainties for this variable.

\section{Topological Susceptibility}
In Ref. \cite{INSTANTON1} we described how to use the scale
invariance of the FP action to define topological charge through
an RG transformation by interpolating the gauge field configuration generated
on a coarse lattice to a finer lattice. The interpolation is done by
 solving Eqn. \ref{STEEP}.
 We then measure the instanton charge on the fine lattice.
 If the configurations are generated using a FP
action and inverse blocked using the same  FP action, the blocking preserves
the topological charge while doubling the radius of any instanton.
In principle this procedure should be repeated until all of the instantons
generated on the coarse lattice have inflated to such a large size that
any measuring technique will give their charge. In practice, we
are limited to a single step of inverse blocking, which might not prove
reliable if the initial configuration is too rough.

We measure the topological charge in the following way: 
We generate
a series of gauge configurations $\{V\}$
on an $L^4$ lattice 
using the eight-parameter approximate
FP action of Table \ref{tab:eightpar}. We then inverse block the
configurations onto a $(2L)^4$ lattice  to produce fine variables
$\{U\}$, by solving  Eqn. \ref{STEEP}.
Finally, we define the topological charge on the configuration as
its value on the fine lattice,
 measured using the geometric definition \cite{GEOMETRIC}
\bee
Q(V) = Q_{geom}(U).
\ee

We use the same  minimizing algorithm as we used for the construction
of the action in Section 2.
Inverse blocking is carried out on the fine lattice and typically takes
50 to 200 minimization passes through the lattice.
We reduced the change in the action per sweep to a value of 0.001
or below, while typical action values range from 1000 to 8000. 
For most configurations this 
accuracy is not necessary, but occasionally the topological charge 
will change as
we reduce the change in the action from, for example, 0.01 to 0.001. For the
topological susceptibility we have not observed any systematic difference, 
though.

To perform a scaling test we compute the dimensionless ratio
of susceptibility  times the fourth power of $r_0$ 
\bee
{ \chi r_0^4} = {\langle Q^2 \rangle}({r_0 \over L})^4 
\ee
where $r_0$ is extracted from the heavy quark potential.
An alternative would be the ratio
of susceptibility divided by the square of the string tension 
\bee
{ \chi \over \sigma^2} = {{\langle Q^2 \rangle} \over {L^4 \sigma^2}}
\ee
but $r_0$ has a smaller overall systematic uncertainty with
respect to form of the potential and fit range.

In measuring the topological charge one encounters two types of
systematic errors even when the action itself shows scaling.
The first is the effect of finite volume: on small volumes larger
instantons are lost and the measured susceptibility is smaller than on
infinite volume. The second effect is related to finite lattice
spacing: small radius instantons are not supported on lattices with
large lattice spacing, again lowering the measured value of the
topological susceptibility. In the following
we compare  the scale invariant quantity 
$\chi r_0^4$ on lattices of fixed physical size, measured
in units of $r_0$  but of different
lattice spacing (again measured in units of $r_0)$.

Table \ref{tab:topol} 
contains our measurements for the topological susceptibility
for the approximate fixed point action.
The data sets typically contain 300-400 independent configurations.
 For comparison we also
present $\langle Q^2 \rangle$ measured directly on the coarse lattice
(no inverse blocking). This number is about a
factor
of 3-4 larger than the theoretically reliable inverse blocked number,
signaling that small dislocations on the coarse lattice have
been incorrectly identified as instantons by the geometric algorithm.
 
For the Wilson action the topological susceptibility has 
been measured  both by the direct geometric definition and 
after cooling (locally minimizing the action)
\cite{COOL}.
Typically, the charge on cooled configurations was found to be about
1/10
the charge measured on coarse configurations.
Cooling has been regarded with misgiving  by some authors, because of
fear that some objects carrying topological charge would disappear
under
cooling. Our results indicate that those fears were justified.
It may still be possible to find a cooling algorithm which reproduces
the results of inverse blocking, but we feel that it would have to be
used with extreme caution. We were not able to invent a satisfactory
 cooling method.

All the results for $\langle Q^2 \rangle_{min}$ of Table \ref{tab:topol} have been obtained after
one inverse blocking step. We performed a short test run at $\beta=1.6$ where we
inverse blocked twice, from $4^4$ to $8^4$ to $16^4$ lattices. On all configurations
the topological charge was identical after one and two inverse blocking steps.

\begin{figure}[htb]
\begin{center}
\leavevmode
\epsfxsize=90mm
\epsfbox{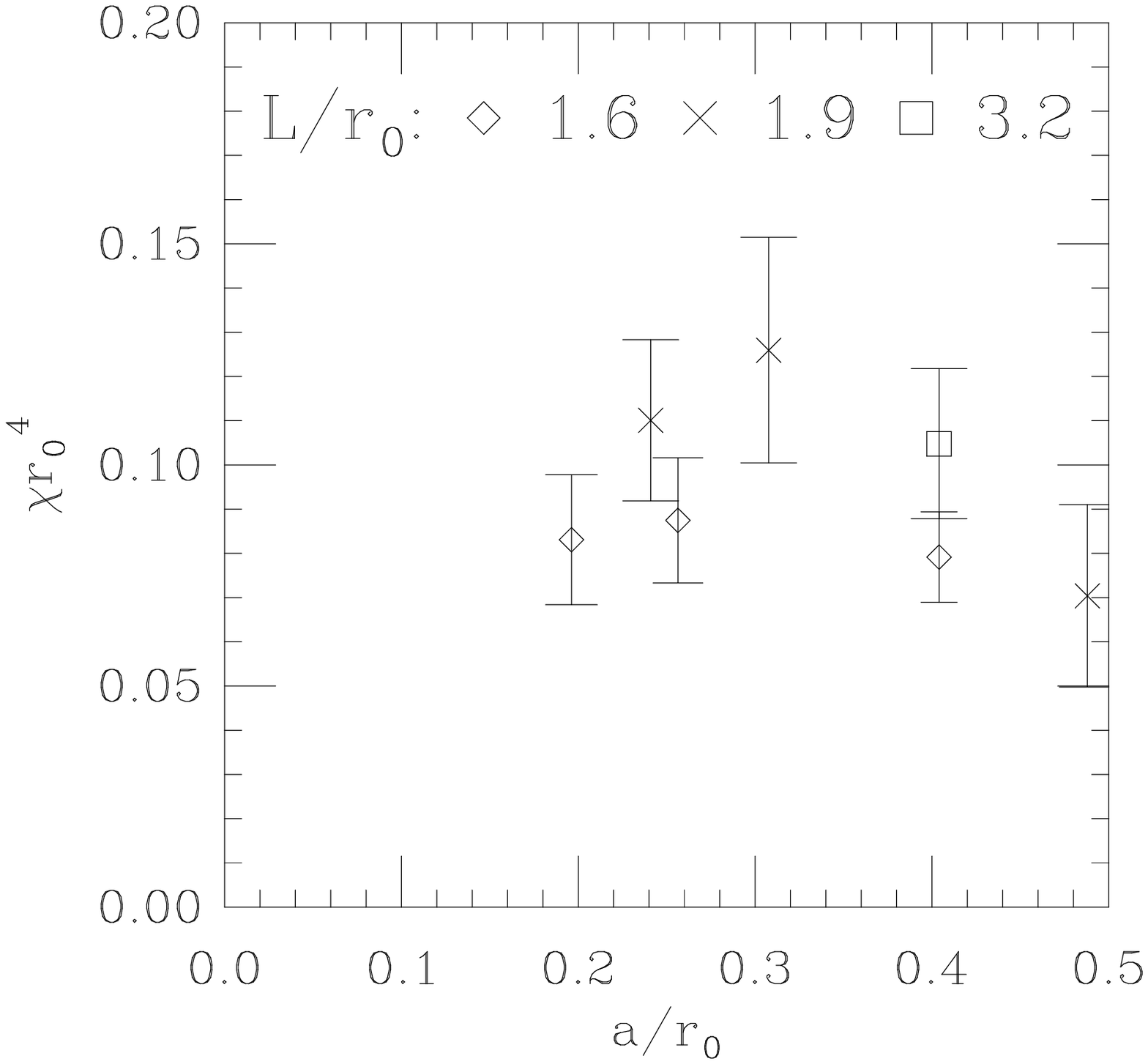}
\end{center}
\caption{
Scaling test for the topological susceptibility,
as a function of lattice spacing. The three lattice sizes (in units of 
$r_0$)
of approximately
1.6, 1.9, and 3.2 are shown by diamonds, crosses, and a square, 
respectively.
}
\label{fig:chir04}
\end{figure}

In Figure \ref{fig:chir04} we plot the quantity $\chi r_0^4$ versus the
lattice spacing $a/r_0$ for physical volumes $L/r_0 \sim 1.6,1.9$ and
$3.2$. The three data points corresponding to  $L/r_0 = 1.54,1.57$ and
$1.62$, which span a range in lattice spacing $a=0.4r_0 - 0.2r_0$ is
consistent with scaling. Apparently a lattice with $a=0.4r_0$ is fine
enough to support the physically relevant instantons. 
To check that assumption further, we created a set of $4^4$
configurations by blocking with the RG transformation of the FP 
a set of $8^4$
configurations generated at $\beta=1.7$. The scaling properties 
and the value of $L/r_0$ of these configurations are the same as the
$8^4$ configurations but the lattice spacing is doubled by the blocking.
 Any change in the topological
susceptibility would be due to the large lattice spacing  $a=0.4r_0$ of the
$4^4$ configurations. 
The blocked lattices are much rougher than the original ones,
and the signal is noisier. It is likely that more than one inverse
blocking step is needed to smooth the configuration.
However, within our accuracy no difference is seen in the topological
susceptibility.

In contrast the
topological susceptibility on the coarsest lattice $a=0.5r_0$ of the
volume $L/r_0=1.85,1.93$ and $1.95$  set is a bit lower then the two 
finer lattice values. This lattice is too coarse for the relevant instantons.
In physical units, a lattice spacing of 0.25 fm is coarse enough to
affect the topological properties of QCD while a spacing of 0.2 fm
seems to be small enough not to do so.
According to our earler work \cite{INSTANTON1}, a candidate instanton solution 
has $Q=1$ only if  $\rho/a > 0.7-0.8$: thus
physically relevant instantons have sizes larger than $0.2\times(0.7-0.8)$
fm or about 0.16 fm.

Within our statistical errors there is no significant change in the
topological susceptibility as we increase the volume of the lattice.

Making the (possibly unwarranted) assumption that the large volume is
large enough to approximate infinity, and making the
(also possibly unwarranted) assumption that the pure gauge SU(2)
$r_0$ is equal to the phenomenological SU(3) number 
0.5 fm, our value $\chi r_0^4= 0.12(2)$ corresponds
to  $\chi$ = ( 235(10) MeV)${}^4$.
Evaluating the Witten-Veneziano formula with the known physical masses of
the appropriate mesons,  and $N_f=3$,
this susceptibility yields an $\eta'$ mass of 1520 MeV.
With the physical $\eta'$ mass,
one would expect to see $\xi^{1/4}=180$ MeV if all the assumptions
which went into the derivation of Eqn. \ref{WZF}
were correct.

\begin{table*}[hbt]
\caption{Topological charge from the approximate FP action.
}
\label{tab:topol}
\begin{tabular*}{\textwidth}{@{}l@{\extracolsep{\fill}}lcccccc}
\hline
$\beta$ &  L & $Q^2$ direct  & $Q^2$ min    &   $L/r_0$ & $\chi r_0^4$ \\
 \hline
1.45  & 4 &  2.79(38) & 1.020(100)  & 1.95   &  0.070(21)  \\
1.502 & 4 &  2.01(8) &  .540(026)  &  1.62 &  0.079(10)  \\
1.502 & 6 &  9.8(8) &  3.15(29)  &  2.42 &  0.091(16)  \\
1.502 & 8 &  37.5(43) &  11.55(87)  &  3.23 &  0.106(17)  \\
1.575 & 6 &  3.96(27) & 1.463(116)  &  1.85 &  0.126(25)  \\
1.6   & 6 &  2.92(12) &  .727(078)  &  1.69  & 0.089(19) \\
1.6  &  8  & 10.2(8) &  3.13(33)    &  2.25 &  0.0140(29)  \\
1.625 & 6 &  1.98(10) &  .490(029)  &  1.54  & 0.087(14)  \\
1.65  & 8  & 5.18(34) & 1.520(105)  &  1.93  & 0.110(18)  \\
1.7  &  8 &  2.21(11) &  .503(030)  &  1.57  & 0.083(15)  \\
\hline
1.7  &  $8\rightarrow 4$ &   &  .60(6)  &  1.57  & 0.099(21)  \\
\hline
\end{tabular*}
\end{table*}

\section{Conclusions}
We demonstrate that an FP action shows scaling in the
ratio $\sigma/T_c$, where the string tension is computed
from the torelon mass,
as opposed to the Wilson action, where the scaling violations are
about eight per cent for the same range of lattice spacings.

We measured the topological susceptibility of SU(2)  pure
gauge theory using a theoretically motivated technique. We found the topological
susceptibility on lattices with $L/r_0 \le 1.6$, $a \le 0.4 r_0$ is
 $\chi r_0^4= 0.12(2)$. In
physical units that corresponds to $\chi^{1/4}=235$ MeV. 

The major bottleneck in the computation is, of course, the inverse blocking.
The inverse block transformation creates a smooth lattice from a  coarse one.
In this sense it is an interpolating
transformation like that of Ref. \cite{HARVARD}. However  
topological properties
are not necessarily preserved by all interpolating transformations. In our
 case Eqn.
\ref{STEEP}  guarantees that the topological charge is the same
 on the coarse lattice and on
the interpolating fine lattice. 
It may be that one can devise an approximate parametrization of the
inverse blocked lattice 
along the lines of the construction for the sigma model described in
Ref. \cite{INSTANTON}, or that
the procedure of 
 Ref. \cite{HARVARD} could be refined for the given RG transformation.
It is also possible that the construction of 
 Ref. \cite{HARVARD} could be used
as a starting point for  further numerical minimization.

The extension of all these methods
to SU(3) appears to be straightforward.  We encourage others interested
in the topological properties of gauge theories to use them.
We remind the reader one last time that the essential ingredient of
our technique is the use of a FP action and of an interpolation 
(inverse blocking) which exploits the scaling properties of the FP
action, to transform lattice configurations which are so rough that
standard measurements of topological charge fail,
into smoother configurations which still retain all the 
original topological properties. On these configurations
the charge can be reliably measured.
The method has to be used with a FP action. Generating configurations
with the Wilson action and inverse blocking them with our RG kernel
in not consistent.

\section{Acknowledgements}
We would like to thank M. M\"uller-Preussker for providing us with 
a copy of the program for measuring topological charge. We very much want to thank Urs Heller for his analysis
of the potential using Wilson loops, described in Section 5.
We would like to thank P.~Hasenfratz and F.~Niedermayer for
useful conversations, and  A. Barker,
M. Horanyi and the Colorado high energy experimental
group for allowing us to use their work stations. 
This work was supported by the U.S. Department of 
Energy and by the National Science Foundation.

\newcommand{\PL}[3]{{Phys. Lett.} {\bf #1} {(19#2)} #3}
\newcommand{\PR}[3]{{Phys. Rev.} {\bf #1} {(19#2)}  #3}
\newcommand{\NP}[3]{{Nucl. Phys.} {\bf #1} {(19#2)} #3}
\newcommand{\PRL}[3]{{Phys. Rev. Lett.} {\bf #1} {(19#2)} #3}
\newcommand{\PREPC}[3]{{Phys. Rep.} {\bf #1} {(19#2)}  #3}
\newcommand{\ZPHYS}[3]{{Z. Phys.} {\bf #1} {(19#2)} #3}
\newcommand{\ANN}[3]{{Ann. Phys. (N.Y.)} {\bf #1} {(19#2)} #3}
\newcommand{\HELV}[3]{{Helv. Phys. Acta} {\bf #1} {(19#2)} #3}
\newcommand{\NC}[3]{{Nuovo Cim.} {\bf #1} {(19#2)} #3}
\newcommand{\CMP}[3]{{Comm. Math. Phys.} {\bf #1} {(19#2)} #3}
\newcommand{\REVMP}[3]{{Rev. Mod. Phys.} {\bf #1} {(19#2)} #3}
\newcommand{\ADD}[3]{{\hspace{.1truecm}}{\bf #1} {(19#2)} #3}
\newcommand{\PA}[3] {{Physica} {\bf #1} {(19#2)} #3}
\newcommand{\JE}[3] {{JETP} {\bf #1} {(19#2)} #3}
\newcommand{\FS}[3] {{Nucl. Phys.} {\bf #1}{[FS#2]} {(19#2)} #3}


\end{document}